\documentclass[12pt]{article}
\usepackage{graphicx}
\usepackage[compress]{cite}
\usepackage{wasysym} 
\usepackage{subfig}

\newcommand\pubnumber{DPF2015-187}
\newcommand\pubdate{August 24, 2015}

\def\iit{Department of Physics\\ Illinois Institute of Technology\\ Chicago,
Illinois 60616-3793, USA}

\def\support{\footnote{Presenting author. kpeders1@hawk.iit.edu}}

\def\Title#1{\begin{center} {\Large #1 } \end{center}}
\def\Author#1{\begin{center}{ \sc #1} \end{center}}
\def\Address#1{\begin{center}{ \it #1} \end{center}}

\newcommand\pubblock{\rightline{\begin{tabular}{l} \pubnumber\\
         \pubdate  \end{tabular}}}

\newenvironment{Presented}{\begin{quotation} \begin{center} 
             PRESENTED AT\end{center}\bigskip 
      \begin{center}\begin{large}}{\end{large}\end{center} \end{quotation}}
\def\Acknowledgments{\bigskip  \bigskip \begin{center} \begin{large}
             \bf ACKNOWLEDGMENTS \end{large}\end{center}}

\def\beq{\begin{equation}}
\def\eeq#1{\label{#1}\end{equation}}
\def\eeqn{\end{equation}}

\def\beqa{\begin{eqnarray}}
\def\eeqa#1{\label{#1}\end{eqnarray}}
\def\eeqan{\end{eqnarray}}

\let\bar=\overbar

\def\Dslash{\not{\hbox{\kern-4pt $D$}}}
\def\dslash{\not{\hbox{\kern-2pt $\del$}}}

\def\msb{{\bar{\ssstyle M \kern -1pt S}}}

\begin{document}
\begin{titlepage}
\pubblock

\vfill
\Title{Flavor tagging TeV jets for BSM and QCD}
\vfill
\Author{Keith Pedersen\support and Zack Sullivan}
\Address{\iit}
\vfill
\begin{abstract}
We present a new scheme for tagging high-$p_{T}$ bottom and charm
jets using energetic muons. Contemporary track-based \emph{b} tags
lose their ability to reject light jet background as jet $p_{T}\rightarrow\mathcal{O}(\mathrm{TeV})$,
where the massive boost exposes fundamental limits in tracking resolution.
For our ``$\mu_{x}$'' tag, the signal efficiency \emph{and} light jet
rejection is robust versus $p_{T}$. In the tested regime (jet $p_{T}\in[\mathrm{0.5,2.1}]$
TeV), $\mu_{x}$ tags $\sim14\%$ of \emph{b} jets,
$\sim6.5\%$ of \emph{c} jets and $\sim0.65\%$
of light jets. Since $\mu_{x}$ tagging should be immediately useful in a
searches for heavy resonances, we test it with a typical dijet
search --- a heavy, leptophobic $Z^{\prime}$.
\end{abstract}
\vfill
\begin{Presented}
DPF 2015\\
The Meeting of the American Physical Society\\
Division of Particles and Fields\\
Ann Arbor, Michigan, August 4--8, 2015\\
\end{Presented}
\vfill
\end{titlepage}
\def\thefootnote{\fnsymbol{footnote}}
\setcounter{footnote}{0}

\section{Introduction}

Searches for physics beyond the standard model (BSM) are a major focus
of research at the LHC. A common signature of BSM physics is a dijet
resonance, which naturally sits atop an enormous QCD background. Perhaps
the most direct way to enhance the purity of such a signal is to flavor tag
the dijets. For instance, isolating the $b\bar{b}$ channel will slash
the dominant light jet background --- provided that the \emph{b} tag
can reject light jets.

Heavy jets (\emph{c} or\emph{ b} initiated) and light jets (\emph{d},
\emph{u}, \emph{s} or \emph{g} initiated) are distinguishable from
their underlying partonic physics. The large mass of heavy quarks
$(m\apprge\Lambda_{QCD})$ discourages fragmentation, so their hadrons
often carry the majority of the jet's momentum. Their proper life-distance
is $\mathcal{O}(0.5\,\mathrm{mm})$, which displaces their decay vertices
a \emph{measurable} distance from primary vertex of the hard scatter,
but at least 50 times closer than more common light hadrons (e.g.
$c\tau(K_{S}^{0})\approx27$ mm). And their large rate of semi-leptonic
decay ($\mathcal{B}(Y_{b/c}\rightarrow l^{+}\nu_{l}X)\approx0.1$
for each $l\in\{e,\mu\}$) enriches their jets with energetic leptons.

These properties are leveraged at the LHC in the two main classes
of flavor tagging.
\begin{itemize}
\item \emph{Track tagging} looks for charged tracks inside a jet
that converge at a secondary vertex (SV) noticeably displaced from
the primary vertex. Using the SV's properties (displacement distance,
reconstructed mass, etc.), jets likely to contain a heavy hadron are
tagged.
\item $p_{T}^{\mathrm{rel}} $\emph{tagging} measures
a lepton's momentum transverse to the centroid of its jet. Heavy jets
contain more leptons, and these should have larger values of $p_{T}^{\mathrm{rel}}$
because: (i) the large mass of its mother causes the lepton to be
emitted at wider angles and (ii) heavy hadrons carry a larger fraction
of the jet's momentum, producing more energetic leptons. However,
$p_{T}^{\mathrm{rel}}$ generally only works for \emph{muons}, as
background electrons are too numerous inside a jet, which is already
an environment where electron identification is difficult.
\end{itemize}
Charm jets perform much weaker in both tags because: (i) \emph{c} hadrons
have shorter lifetimes and smaller masses and (ii) \emph{b} hadrons
primarily decay to \emph{c} hadrons, giving \emph{b} jets twice as
many muons and a second chance to create displaced tracks. Thus, both
tags are generally considered \emph{b} tags with a higher fake
rate for \emph{c} jets than light jets.

As jet $p_{T}\rightarrow\mathcal{O}(\mathrm{TeV})$, both tags
lose much of their ability to reject light jet background. Here, the
extreme boost collimates the jet so much that SV properties become
very sensitive to tracking resolution \cite{ATLAS:7Tev-bTag-Commissioning, CMS:8Tev-bTag-Performance}.
Similarly, $p_{T}^{\mathrm{rel}}$ distributions for muons in heavy and light jets 
become nearly indistinguishable \cite{ATLAS:Calibrate-bTag-ptRel}.
However, while the loss of purity for the track tag is primarily a
detector effect, the failure of $p_{T}^{\mathrm{rel}}$ is predictable
from the underlying kinematics of boosted semi-muonic decay.

\section{A new heavy flavor tag}

\begin{figure}[h]
\begin{centering}
\includegraphics[width=0.80\textwidth]{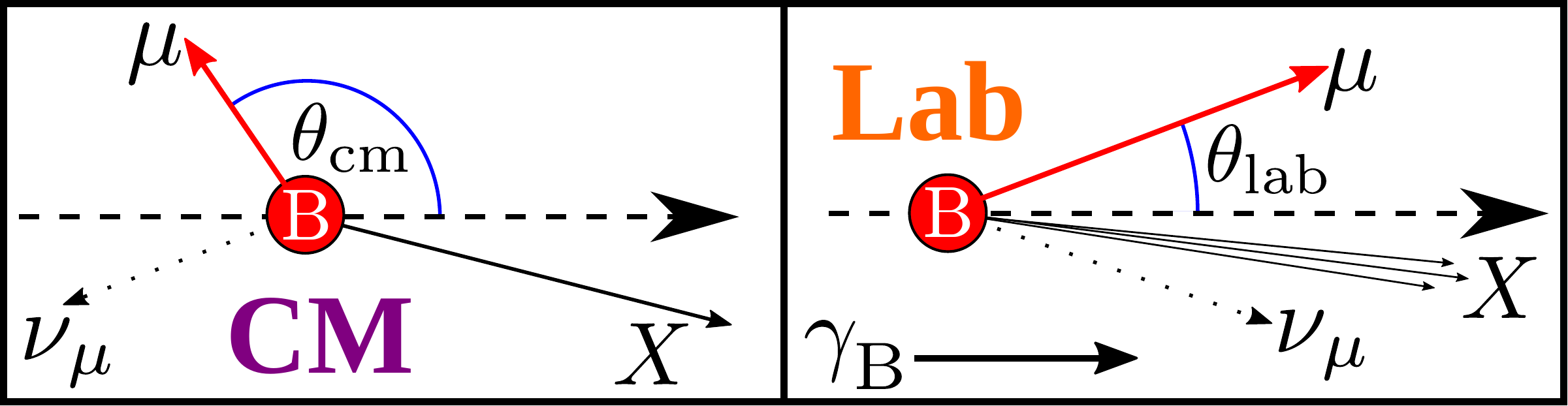}
\caption{Boosted nomenclature.\label{fig:Boosted-Nomenclature}}
\end{centering}
\end{figure}

Consider a jet containing a \emph{B} hadron. In the center-of-momentum
(CM) frame, a muon is emitted with some speed $\beta_{\mu,\mathrm{cm}}$
and at some angle $\theta_{\mathrm{cm}}$ w.r.t. the boost axis (see
Fig. \ref{fig:Boosted-Nomenclature}). In the lab frame, the boost
$\gamma_{B}^{\,}$ compresses the decay products into a subjet. 
Using $\kappa\equiv\beta_{\mathrm{B}}/\beta_{\mu,\mathrm{cm}}$,
we can define a lab frame observable
\begin{equation}
x\equiv\gamma_{\mathrm{B}}^{\,}\,\tan(\theta_{\mathrm{lab}})=\frac{\sin(\theta_{\mathrm{cm}})}{\kappa+\cos(\theta_{\mathrm{cm}})}\quad.\label{eq: x definition}
\end{equation}
When the muon and the \emph{b} hadron are both relativistic, $\kappa\approx1$
and $x$ is nearly invariant. Since we are only interested in \emph{b} jets
where $\gamma_{\mathrm{B}}^{\,}\gg\gamma_{\mu,\mathrm{cm}}^{\,}$,
we only consider the \emph{over}-boosted $(\kappa\ge1)$ distribution
for count $N$ (where $x\in[0,1/\sqrt{\kappa^{2}-1}]$),
\begin{equation}
\frac{dN}{dx}=4\pi\,\frac{2x}{(x^{2}+1)^{2}}\,K(x,\kappa)\quad\textrm{with}\label{eq:dN_dx}
\end{equation}
\begin{equation}
K(x,\kappa)=\frac{(1+\kappa^{2})+x^{2}(1-\kappa^{2})}{2\sqrt{1+x^{2}(1-\kappa^{2})}}\quad.
\end{equation}
Here, $K(x,\kappa)$ corrects the nominal shape when $\kappa>1$.

\begin{figure}[h]
\begin{centering}
     \subfloat[]{\includegraphics[width=.49\textwidth]{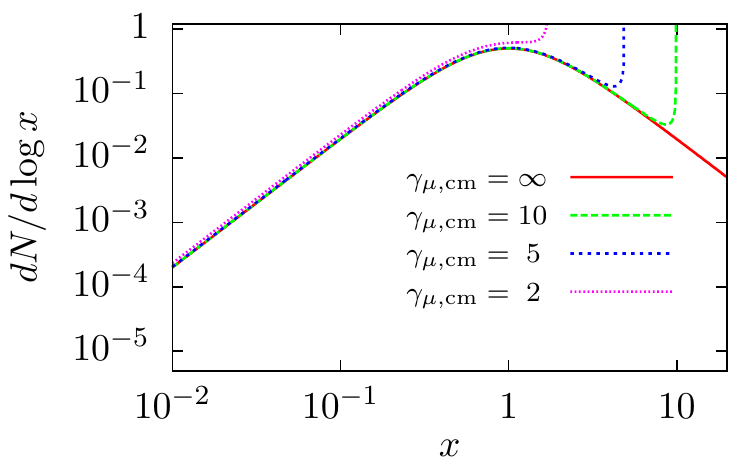}\label{fig:theo-dNdLogx}}
     \subfloat[]{\includegraphics[width=.49\textwidth]{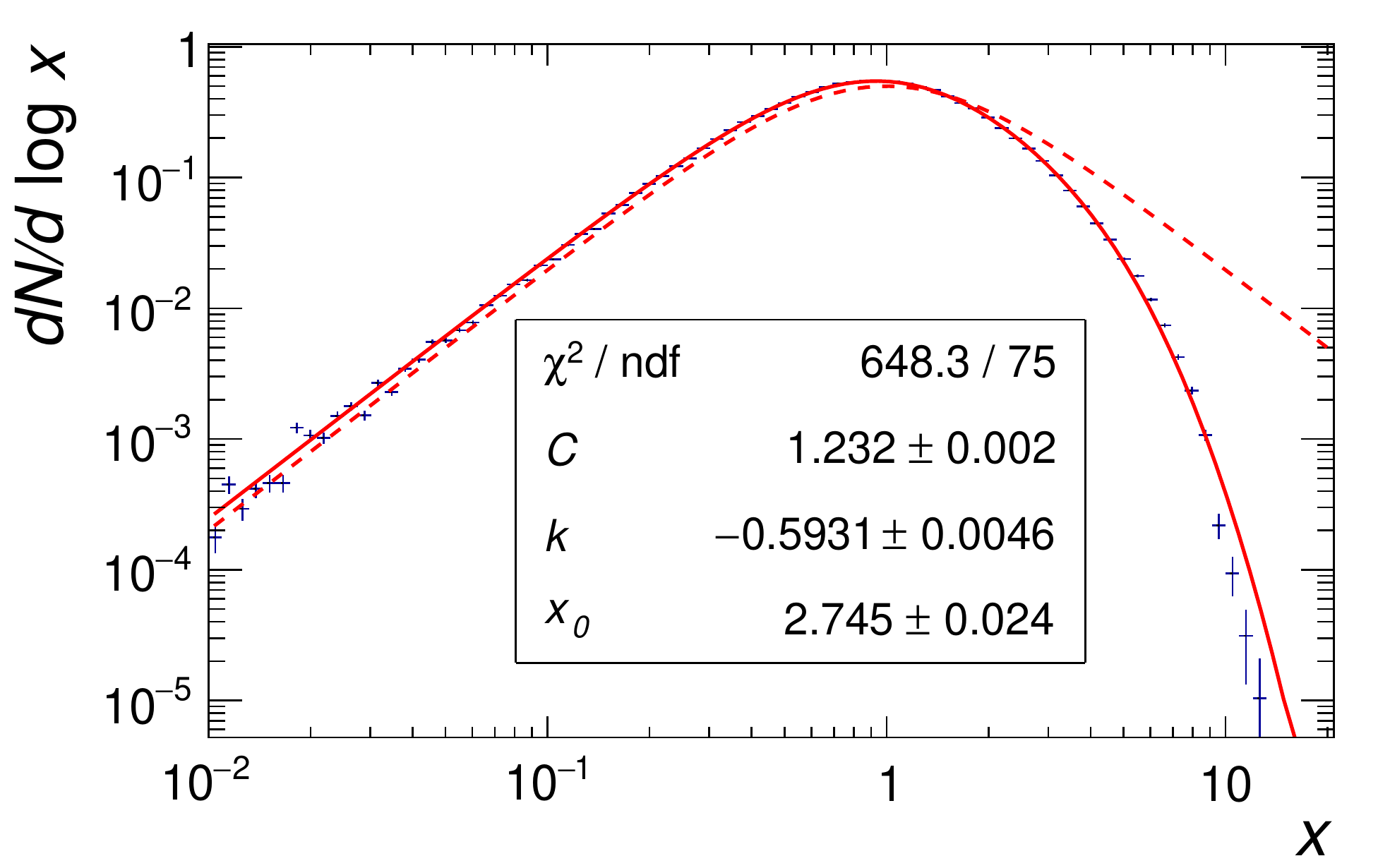}\label{fig:bjet-dNdLogx}}
     \caption{(a) Theoretical $dN/d\log{x}$ for muons of various $\gamma_{\mathrm{\mu},\mathrm{cm}}^{\,}$ and (b) $dN/d\log x$ for muons inside simulated \emph{b} jets.}
\end{centering}
\end{figure}

Figure \ref{fig:theo-dNdLogx} demonstrates that the $K(x,\kappa)$
correction is small for most $\gamma_{\mathrm{\mu},\mathrm{cm}}^{\,}$.
This is further exemplified in Fig. \ref{fig:bjet-dNdLogx}, where
an ensemble of decays ``integrates'' over the $\gamma_{\mu,cm}$
spectrum, while essentially preserving the nominal shape ($\kappa=1$,
dotted line) of Eq. (\ref{eq:dN_dx}). The downward correction
in the tail, from each muon's boost cone boundary, is fit by adding
a Logistic curve to the nominal shape (to act as a cut-off function),
along with a normalization factor $C$.
%
%
%
%

These results support the calculation that, if most muons are relativistic
in the CM frame, at least 90\% will arrive in a cone defined by $x\le3$.
This cone can be used to accept/exclude muons which are consistent
with boosted semi-muonic decay, forming the basis for a new \emph{b} tag.
Most notably, the \emph{physical} size of this cone should decrease
as jet $p_{T}$ increases, underlying the failure of $p_{T}^{rel}$;
muons from \emph{very} boosted \emph{b} hadrons should no longer
arrive at wide angles.

\subsection{Measuring $x$}

Measuring $x$ requires reconstructing the four-momentum of the semi-muonic
decay,
\begin{equation}
\mathbf{p}_{\mathrm{subjet}}=\mathbf{p}_{\mathrm{core}}+\mathbf{p}_{\mu}+\mathbf{p}_{\nu_{\mu}}\approx\mathbf{p}_{\mathrm{core}}+2\mathbf{p}_{\mu}\quad,
\end{equation}
where the ``core'' is composed of the boosted hadronic remnants
(the $X$ in Fig. \ref{fig:Boosted-Nomenclature}). Since most of
the $\nu_{\mu}$'s lab frame momentum is from its mother's boost,
the muon is an acceptable proxy (with the simplest choice being $\mathbf{p}_{v_{\mu}}=\mathbf{p}_{\mu}$).

Tracks provide the best angular information to reconstruct the thin core,
but its high collimation will hamper track finding in a non-trivial way. 
To simplify detector simulation, we build jets (and
cores) solely from calorimeter towers and muons. To mitigate the coarse
granularity of the hadronic calorimeter (HCal), we use the finer granularity
of the the EM Calorimeter (ECal) to orient the combined towers (``ECaL
pointing''). This assumes that cores likely contain boosted
photons/electrons (especially from $\pi^{0}$), and most hadrons begin
showering in the ECal. 

Jets containing muons are reclustered with anti-$k_{T}$ to produces a list of candidate cores.
The fixed tower width $w$ requires using $\sqrt{2}w<R_{\mathrm{core}}<2w$ 
to capture the core in a $3\times3$
grid (the smallest choice which can triangulate an impact). Since
this granularity produces an ill-measured mass, we fix each candidate
to $m_{\mathrm{core}}$ (the expected mass under the \emph{b} hadron
hypothesis). The ``correct'' core is the one which, when the muon
is added twice, produces a subjet whose mass is closest to $m_{\mathrm{B}}$,
the nominal mass of the \emph{b} hadron admixture. As a final sanity test, 
the momentum fraction of the subjet
\begin{equation}
f_{\mathrm{subjet}}=\frac{p_{T,\mathrm{subjet}}}{p_{T,\mathrm{jet}}}\quad,
\end{equation}
should be close to one, since \emph{b} quarks shun hard radiation.

\subsection{The $\mu_{x}$ tag}

The $\mu_{x}$ tag uses four basic cuts: (i) $p_{T,\mu}\geq10$ GeV
ensures that the muon is well reconstructed, (ii) jet $p_{T}\geq300$
GeV confirms that boosted kinematics apply, (iii) $x\leq3$ establishes
that the muon is consistent with a boosted primordial decay, given
its local jet environment and (iv) $f_{\mathrm{subjet}}\geq0.5$ verifies
that the subjet is consistent with a heavy quark.

We test the $\mu_{x}$ tag by generating samples of
$b\bar{b}$, $c\bar{c}$, and $j\bar{j}$ (where $j\in\{u,d,s,g\}$),
spanning $p_{T}=$ 0.1--2.1 TeV, with {\sc MadGraph}5 \cite{MadGraph:5}.
We fragment and hadronize in {\sc Pythia} 8 \cite{Pythia:Physics, Pythia:8.1}
and model the ATLAS detector with {\sc Delphes} 3 \cite{Delphes:3}, clustering jets
with {\tt FastJet} 3 \cite{FastJet:User-Manual}. Pileup is generated by {\sc Pythia}, 
using the parameters suggested by ATLAS in Ref. \cite{ATLAS:Pileup-2011}, 
and a random number of pileup events (drawn from
a Poisson distribution with $\mu=40$) are added to each event.

Muons are simulated as ``standalone'' tracks, which only use hits
from the ATLAS Muon Spectrometer (MS). In order to implement ``ECal pointing''
in the {\sc Delphes} Calorimeter module, we use the granularity of ATLAS ECal Layer
2 ($0.025\times0.025$) for the region overlapping the MS ($\left|\eta\right|<2.7$).
Jets are clustered from towers and muons using anti-$k_{T}$ with
$R=0.4$. The subjets of those containing muons are built using $R_{\mathrm{core}}=0.04$,
$m_{\mathrm{core}}=2$ GeV, and $m_{\mathrm{B}}=5.3$ GeV. To prevent
$x$ from growing absurdly small, we use $\gamma_{\mathrm{B}}^{\,}=E_{\mathrm{subjet}}/\min(m_{\mathrm{subjet}},\textrm{12 GeV})$. 
\begin{figure}[h]
\subfloat[]{\includegraphics[width=0.45\textwidth]{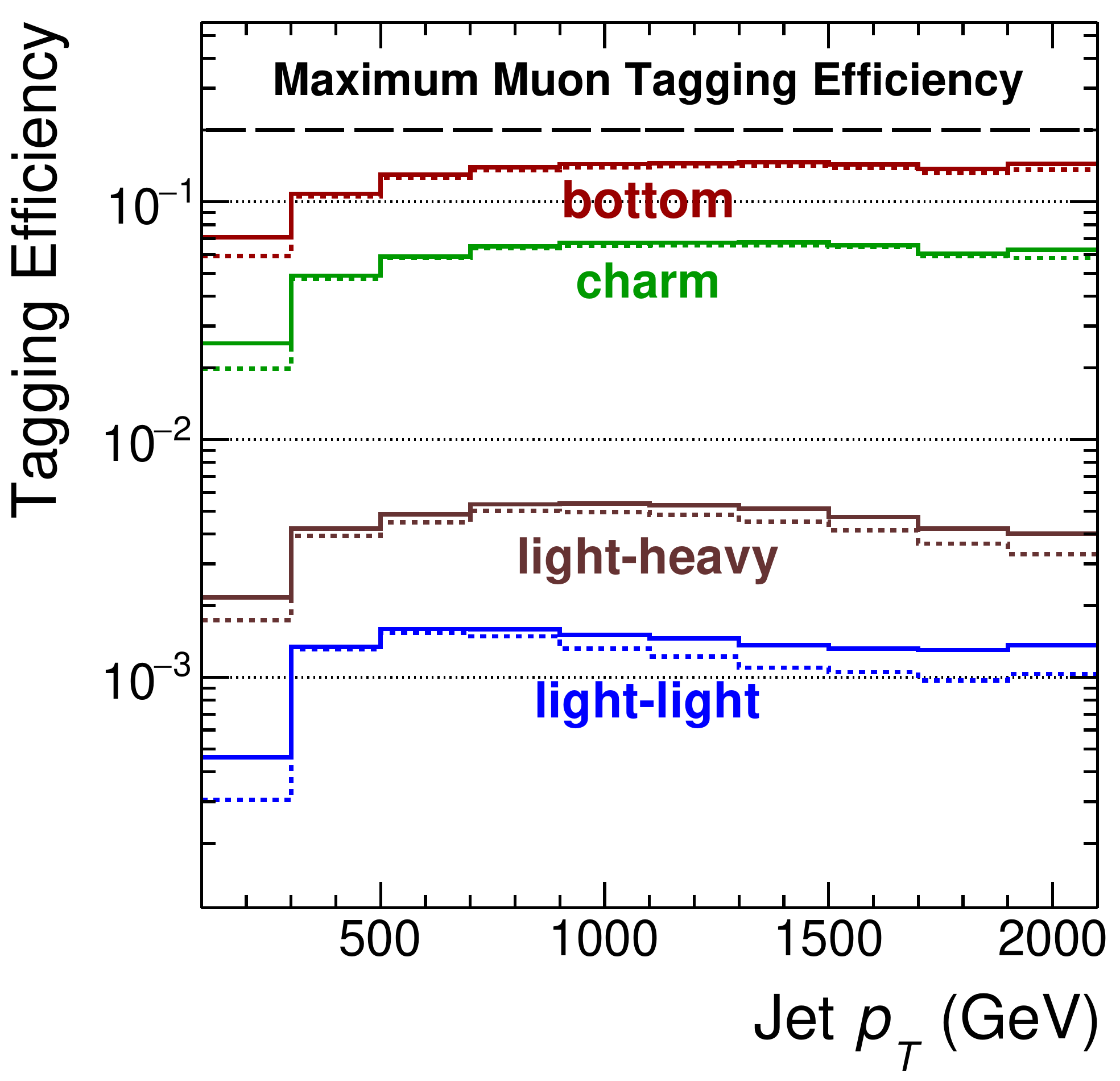}}
\quad\quad
\subfloat[]{\includegraphics[width=0.45\textwidth]{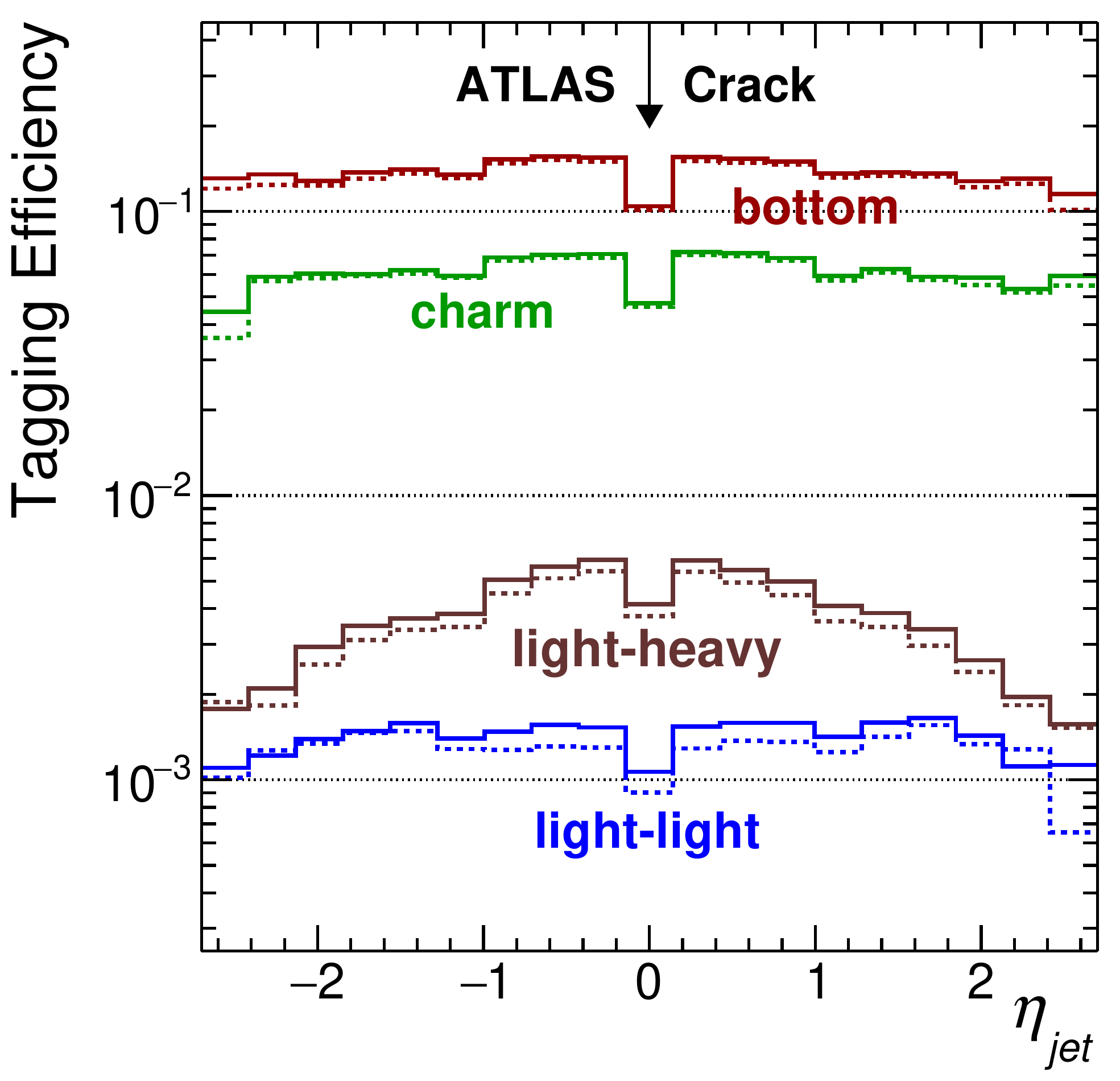}\label{fig:efficiency-versus-eta}}
\caption{$\mu_{x}$ tagging efficiency without pileup (solid) and with $\mu=40$ (dashed) versus (a) jet $p_{T}$ and (b) $\eta_{jet}$.}\label{fig:Tagging-Efficiency}
\end{figure}

In Fig. \ref{fig:Tagging-Efficiency} we show the efficiency to tag
the top two jets (ranked by $p_{T}$) in each event. Light jets are
split into two classes, where \emph{light-heavy} is a jet initiated
by a light parton, but whose muon originated from a heavy hadron (i.e.
a gluon split to heavy quarks during fragmentation). Once the boosted
approximations turn on ($\kappa\approx1$), the efficiency to tag
heavy jets is both flat versus $p_{T}$ and insensitive to pileup.
Fig. \ref{fig:efficiency-versus-eta} only uses jets with $p_{T}\geq300$
GeV, leading to poor statistics at the edge of the MS; nonetheless,
the efficiency to tag heavy jets is also relatively flat with $\eta_{jet}$
(excluding ATLAS's central detector services crack, where standalone
muon efficiency plummets).

\section{Leptophobic \textmd{\textup{\normalsize{}$Z^{\prime}$}}}

One of the simplest BSM models is an additional $U(1)^{\prime}$ symmetry
mediated by a heavy, neutral $Z^{\prime}$. Since the LHC has already
probed large scales in the ``golden'' channel (dileptons) and seen
no such resonance, the simplest $U(1)^{\prime}$ are excluded. But
if the new symmetry is associated with baryon number, only SM quarks
would be charged. Anomaly cancellation requires this $U(1)_{B}^{\prime}$
to come with new, vector-like quarks and at least one scalar field
whose VEV breaks the symmetry \cite{Dobrescu:2013coa, Dobrescu:2015asa}.
If the vector-like quarks are kinematically inaccessible at the LHC, then the
$Z_{B\mu}^{\prime}$ gauge coupling to SM quarks \cite{Dobrescu:2013coa}
\begin{equation}
\frac{g_{B}}{6}Z_{B\mu}^{\prime}\bar{q}\gamma^{\mu}q\quad,
\end{equation}
would be the only experimental signature of the $U(1)_{B}^{\prime}$
at leading order. Since $\mu_{x}$ tagging can greatly enhance the purity
of such a dijet signal, we simulate a ``bump hunt'' at ATLAS Run
II (i.e. looking for an excess in $d\sigma/dm_{jj}$). We use two
classes: 1-tag and 2-tag (where $N$-tag requires tagging at least
$N$ of the top two jets, ranked by $p_{T}$).

The relevant signal is $pp\rightarrow Z_{B}^{\prime}\rightarrow b\overline{b}/c\bar{c}(+j)$,
where the optional light jet radiation enhances the signal cross section.
The relevant background for each class is purely QCD; yet while both
classes use $pp\rightarrow b\overline{b}/c\bar{c}/jj(+j)$, the 1-tag
class also draws heavily from the bottom/charm PDF via $jh\rightarrow jh(+j)$
(a heavy quark scattering from a light parton). We generate all samples
at $\sqrt{s}=13$ TeV using CT14llo PDFs \cite{CTEQ:CT14}, improving their differential
cross sections by using MLM matching between MadGraph and Pythia (in
``shower-kt'' mode with $q_{cut}\approx M_{Z_{B}^{\prime}}/20$)\cite{Alwall:2008qv}.
Signal sets are generated for a range of $M_{Z_{B}^{\prime}}$, with
corresponding background sets governed by identical kinematic/matching
cuts. 

Because the $\mu_{x}$ light jet efficiency is minuscule, we approximate
the second tag for the light dijet background. This is accomplished
by fitting the light jet efficiency as a function of jet $p_{T}$
and $\eta$. When exactly one leading jet is tagged, we find the probability
to tag the other jet, then re-weight the event by a factor of $A_{1}=\frac{1-\rho/2}{1-\rho}$
or $A_{2}=\frac{\rho}{2(1-\rho)}$, for the 1-tag and 2-tag classes,
respectively. Light dijet events with two real tags are discarded,
to prevent double-counting.

We require $\Delta R_{jj}\le1.5$, to suppress $t$-channel background,
and insist that both jets fall withing the MS ($\left|\eta_{jet}\right|\le2.7$).
Even though we generate an optional radiation jet, we find that adding
a hard third jet to the tagged dijet system causes an unacceptable
hardening of the steeply falling QCD continuum. The subsequent
loss of the radiation jet, combined with the neutrino estimation inherent
to $\mu_{x}$ tagging, smears the dijet mass, requiring a rather wide
mass window $(\left[0.85,1.25\right]\times M_{Z^{\prime}})$ designed
to capture nearly all signal above $M_{Z^{\prime}}$ and as much
signal below, while only doubling the background.

\begin{figure}[h]
\begin{centering}
\includegraphics[width=0.90\textwidth]{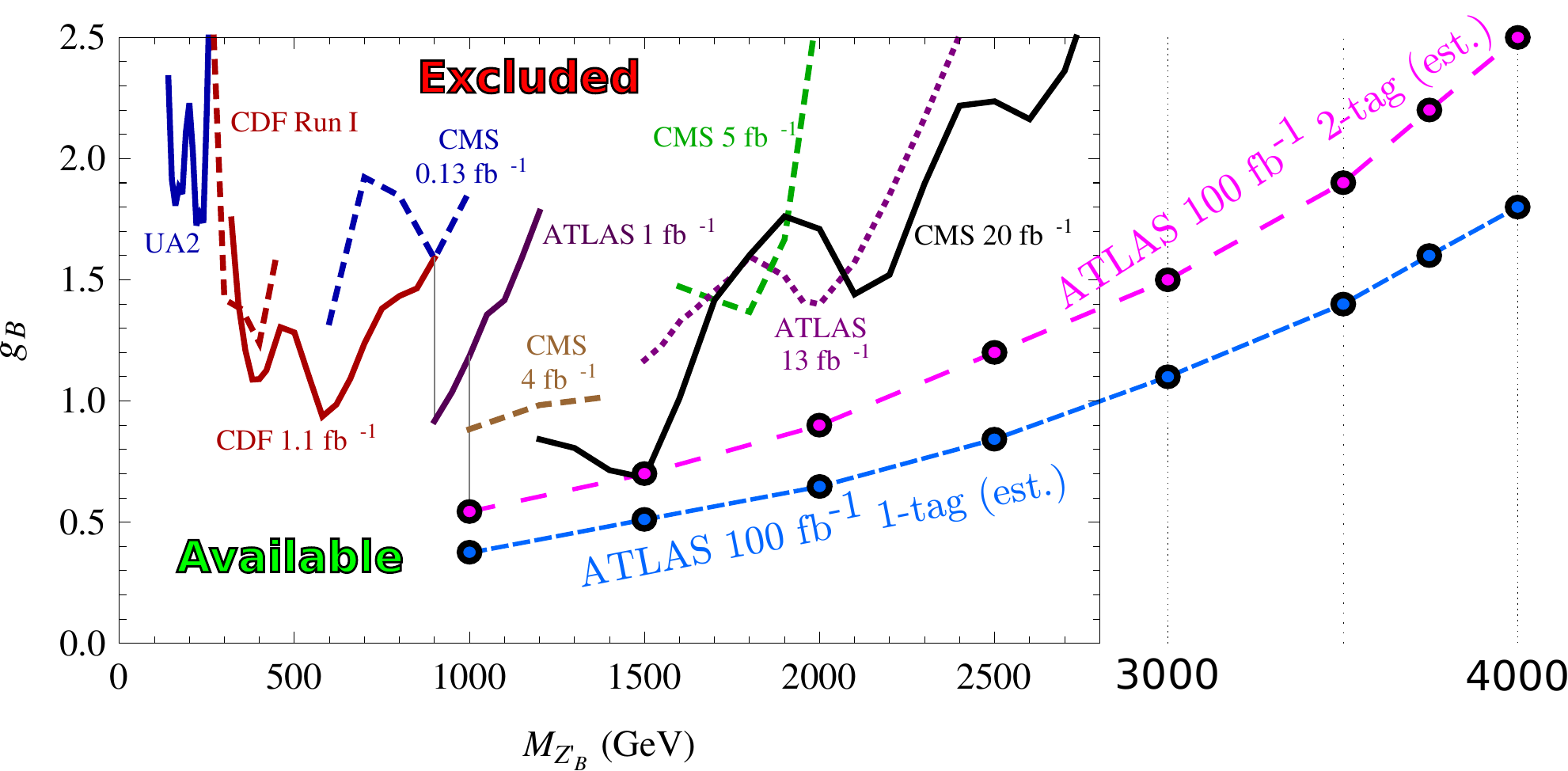}
\caption{95\% confidence level exclusion limits for $Z_{B}^{\prime}$ models.}\label{fig:Exclusion-Limits}
\end{centering}
\end{figure}

We compare the experimental
reach of the $\mu_{x}$ tag to existing dijet searches
via an exclusion plot  \cite{Dobrescu:2013coa} for a leptophobic $Z_{B}^{\prime}$ (Fig. \ref{fig:Exclusion-Limits}),
which demonstrates that --- given an expected Run II luminosity of
100 $\mathrm{fb}^{-1}$ --- the $\mu_{x}$ tag should be sensitive
to entirely new regions of model-space.

\section{Conclusion}

The $\mu_{x}$ tag is a high $p_{T}$, heavy flavor tag whose signal
efficiency \emph{and} light jet rejection are robust versus $p_{T}$.
It performs well at identifying a generic heavy quark signal, which
suggests it will be useful in a range of applications in BSM physics
(especially models which couple predominantly to heavy flavors). Additionally,
since $\mu_{x}$ and track tagging are not mutually exclusive, they
should be able to cross-check one another in the high $p_{T}$ regime
(a region where track tags are dominated by uncertainties in their
tagging efficiency). This should create a combined flavor tag with
an overall higher efficiency, a tunable light jet rejection, and much
lower uncertainties in its high-$p_{T}$ tagging efficiency.

\Acknowledgments
This work was supported by the U.S.\ Department of Energy under Award
No.\ {DE-SC0008347}.

\end{document}